\documentclass[10pt,twoside,twocolumn,english,aps,manuscript,preprint,showpacs]{revtex4}
\usepackage[T1]{fontenc}
\usepackage[latin9]{inputenc}
\usepackage{color}
\usepackage{textcomp}
\usepackage{amstext}
\usepackage{graphicx}

\makeatletter

\newcommand{\lyxmathsym}[1]{\ifmmode\begingroup\def\b@ld{bold}
  \text{\ifx\math@version\b@ld\bfseries\fi#1}\endgroup\else#1\fi}

\providecommand{\tabularnewline}{\\}

\@ifundefined{textcolor}{}
{%
 \definecolor{BLACK}{gray}{0}
 \definecolor{WHITE}{gray}{1}
 \definecolor{RED}{rgb}{1,0,0}
 \definecolor{GREEN}{rgb}{0,1,0}
 \definecolor{BLUE}{rgb}{0,0,1}
 \definecolor{CYAN}{cmyk}{1,0,0,0}
 \definecolor{MAGENTA}{cmyk}{0,1,0,0}
 \definecolor{YELLOW}{cmyk}{0,0,1,0}
}

\makeatother

\makeatother

\usepackage{babel}
\begin{document}

\title{Frustration induced disordered magnetism in Ba$_{3}$RuTi$_{2}$O$_{9}$}

\author{Tusharkanti Dey}

\email[Email: ]{tusdey@gmail.com}

\affiliation{Department of Physics, Indian Institute of Technology Bombay, Powai,
Mumbai 400076, India }

\author{A.V. Mahajan}

\affiliation{Department of Physics, Indian Institute of Technology Bombay, Powai,
Mumbai 400076, India }
\begin{abstract}
The title compound Ba$_{3}$RuTi$_{2}$O$_{9}$ crystallizes with
a hexagonal unit cell. It contains layers of edge shared triangular
network of Ru$^{4+}$($S=1$) ions. Magnetic susceptibility $\chi(T)$
and heat capacity data show no long range magnetic ordering down to
$1.8$\,K. A Curie-Weiss (CW) fitting of $\chi(T)$ yields a large
antiferromagnetic CW temperature $\theta_{\mathrm{CW}}=-166$\,K.
However, in low field, a splitting of zero field cooled (ZFC) and
field cooled (FC) $\chi(T)$ is observed below $\sim30$\,K. Our
measurements suggest that Ba$_{3}$RuTi$_{2}$O$_{9}$ is a highly
frustrated system but only a small fraction of the spins in this system
undergo a transition to a frozen magnetic state below $\sim30$\,K.
\end{abstract}

\pacs{75.47.Lx, 75.50.Lk, 75.40.Cx}

\maketitle

\section{introduction}

In recent times, $4d$ and $5d$ transition metal based materials
have been in the focus as their driving physics is different from
their $3d$ counterparts but have not been explored much. This variation
arises because of their extended `$d$' orbitals, a small onsite Coulomb
energy `U', and a large spin-orbit coupling (SOC) \cite{Kim-PRL-101-2008}.
The interplay between U and SOC makes these materials interesting.
Many systems of the $4d/5d$ family like Sr$_{2}$(Ir/Ru/Rh)O$_{4}$
\cite{Kim-PRL-101-2008,Haverkort-PRL-2008-Sr2Ru-RhO4,Kim-Science-2009},
Ca$_{2}$RuO$_{4}$ \cite{Mizokawa-PRL-87-2001-Ca2RuO4}, Na$_{2}$IrO$_{3}$
\cite{Singh-PRB-82-2010}, Na$_{4}$Ir$_{3}$O$_{8}$ \cite{Okamoto(Na4Ir3O8)-PRL-99-2007},
(Sr/Ca)$_{3}$Ru$_{2}$O$_{7}$ \cite{Perry(Sr3Ru2O7)-PRL-86-2001,Cao-NJP-2004-Ca3Ru2O7}
etc. have drawn attention in recent years.

Recently, we reported results of our investigations of hexagonal Ba$_{3}$IrTi$_{2}$O$_{9}$
with $5d$ Ir$^{4+}$ (possibly$J_{eff}=1/2$) ions forming a $2$D
triangular network \cite{Dey-PRB-86-2012}. A large site-disorder
between Ir$^{4+}$ and Ti$^{4+}$ cations plays an important role
in the magnetic properties of the sample. The system remains paramagnetic
down to $0.35$\,K despite a large antiferromagnetic (AF) Curie-Weiss
temperature $\theta_{\mathrm{CW}}\sim-130$\,K. Magnetic heat capacity
at low temperature follows a power law with temperature. These results
suggested that Ba$_{3}$IrTi$_{2}$O$_{9}$ is in a spin liquid ground
state with $5d$ Ir$^{4+}$ (possibly $J_{eff}=1/2$) in a $2$D triangular
network. The SOC presumably plays a significant role in determining
the properties of the system. It will be interesting to study its
$4d$-analog (say, based on Ru) where spin-orbit coupling effect is
expected to be of intermediate strength compared to that in $3d$
and $5d$ elements. 

Dickson\textit{ et al}. \cite{Dickson-JACS-1961} measured the magnetic
susceptibility variation with temperature of hexagonal Ba$_{3}$RuTi$_{2}$O$_{9}$
in the limited temperature range $253$\,K to $323$\,K. Susceptibility
data in a larger temperature range $77-335$\,K are reported in Ref.
\cite{Byrne-JSSC-2-1970}. In both the papers, a high, AF $\theta_{\mathrm{CW}}$
($\sim-250$\,K) is inferred \cite{Byrne-JSSC-2-1970,Dickson-JACS-1961}.
These reports indicate that Ba$_{3}$RuTi$_{2}$O$_{9}$ could also
be a frustrated spin liquid candidate and a $4d$ analog of Ba$_{3}$IrTi$_{2}$O$_{9}$
with $S=1$. Thus low temperature measurements on Ba$_{3}$RuTi$_{2}$O$_{9}$
are warranted to confirm its magnetic ground state.

In this paper we report preparation, structural analysis, magnetic
susceptibility and heat capacity measurements on Ba$_{3}$RuTi$_{2}$O$_{9}$.
The system crystallizes in the space group P6$_{3}$mc with a large
site sharing between Ru$^{4+}$ and Ti$^{4+}$ ions as reported earlier
\cite{Radtke-PRB-81-2010} and similar to that in  Ba$_{3}$IrTi$_{2}$O$_{9}$
\cite{Dey-PRB-86-2012}. Susceptibility data show no long range ordering
down to $2$\,K but splitting between zero field cooled (ZFC) and
field cooled (FC) data is found below $30$\,K. A large $\theta_{\mathrm{CW}}$
(AF) obtained from CW fitting indicates strong correlation between
the magnetic ions. No anomaly, even at $30$\,K, is found in heat
capacity measurements down to $1.8$\,K. These measurements suggest
that the system is highly frustrated and perhaps a fraction of the
spins form a disordered magnetic state below $30$\,K.

\section{experimental details}

A polycrystalline sample of Ba$_{3}$RuTi$_{2}$O$_{9}$ was prepared
by conventional solid state reaction method using high purity BaCO$_{3}$,
TiO$_{2}$ and Ru metal powder as starting materials. Stoichiometric
amount of the starting materials were mixed thoroughly, pressed into
a pellet and calcined at $900^{0}$C for $15$\,h. After calcination,
the pellet was crushed into powder, pelletized and fired at $1100^{0}$C
for $100$\,h with several intermediate grindings.

Powder x-ray diffraction (XRD) measurements were performed at room
temperature with Cu $K_{\alpha}$ radiation ($\lambda=1.54182\textrm{\AA}$)
in a PANalytical X\textquoteright{}Pert PRO diffractometer using Si
for calibration. Magnetization $M$ measurements were carried out
in the temperature $T$ range $2-400$\,K using a Quantum Design
PPMS and SQUID VSM. Heat capacity measurements were done in the temperature
range $1.8-300$\,K and field $H$ range $0-9$\,T using the heat
capacity attachment of Quantum Design PPMS.

\section{results and discussion}

\begin{figure}
\centering{}\includegraphics[scale=0.7]{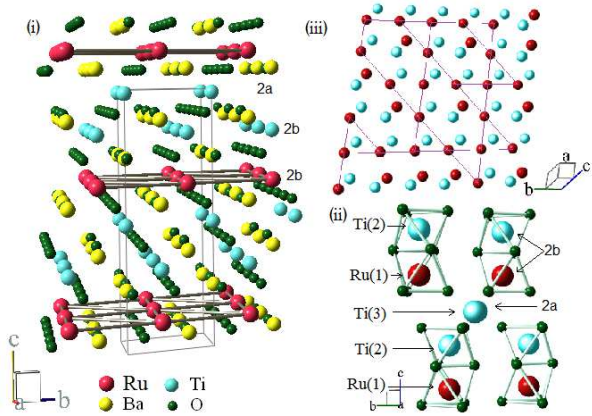}\caption{\label{fig:Structure} (i) The ideal structure of Ba$_{3}$RuTi$_{2}$O$_{9}$
without any site disorder between Ti$^{4+}$ and Ru$^{4+}$ ions is
shown. The Ru$^{4+}$ spins form an edge shared triangular network
parallel to the $ab$ plane as shown. (ii) The RuTiO$_{9}$ bioctahedra
are shown (iii) A possible arrangement of Ti$^{4+}$ and Ru$^{4+}$
ions in the $ab$ plane is shown when about $33\%$ of Ru$^{4+}$
ions exchange their positions with Ti$^{4+}$ ions in the $2$b site. }
\end{figure}

Single phase nature of our sample was confirmed from the XRD measurements.
Maunders $et$ $al$. have reported Ba$_{3}$RuTi$_{2}$O$_{9}$ to
crystallize in the space group P6$_{3}mc$ \cite{Maunders-ActaCrysta-61-2005}.
In this space group, the $2$a site is occupied by the Ti$^{4+}$
ions and in the $2$b sites Ti$^{4+}$/Ru$^{4+}$ ions reside in an
ordered manner. In the ideal case, the ordered arrangement of Ti$^{4+}$
and Ru$^{4+}$ ions form face-sharing RuTiO$_{9}$ bioctahedra (Fig.
\ref{fig:Structure}(ii)) and the Ru$^{4+}$ spins form an edge-shared
triangular network in the $ab$ plane (as shown in Fig. \ref{fig:Structure}(i)).
However, a site sharing between Ti$^{4+}$ and Ru$^{4+}$ ions is
expected as the two ions are of similar ionic size. In Ref. \cite{Radtke-PRB-81-2010}
four possible arrangements of Ru$^{4+}$/Ti$^{4+}$ ions have been
suggested in Ba$_{3}$RuTi$_{2}$O$_{9}$. Among them, the two arrangements
where RuTiO$_{9}$ and Ru$_{2}$O$_{9}$/Ti$_{2}$O$_{9}$ bioctahedra
are formed have been found to be energetically more probable \cite{Radtke-PRB-81-2010}.
We refined our XRD data using the space group P6$_{3}mc$ and considering
a site sharing between Ti$^{4+}$ and Ru$^{4+}$ ions. We have found
almost $31\%$ site exchange of Ru$^{4+}$ ions with Ti$^{4+}$ ions
in the $2$b site and an $8\%$ site sharing with $2$a site Ti$^{4+}$
ions. This situation is very similar to that found in Ref. \cite{Radtke-PRB-81-2010}.
In Ba$_{3}$IrTi$_{2}$O$_{9}$ as well, we found a large site sharing
of Ir$^{4+}$ ions with Ti$^{4+}$ ions both in $2$b as well as $2$a
sites \cite{Dey-PRB-86-2012}.

\begin{figure}
\centering{}\includegraphics[scale=0.3]{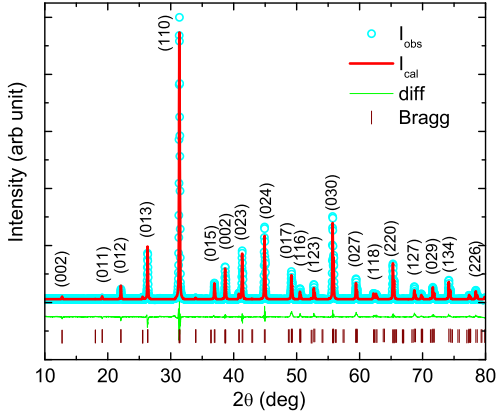}\caption{\label{fig:XRD} The x-ray diffraction pattern of Ba$_{3}$RuTi$_{2}$O$_{9}$
is shown. The refinement of these data using the space group P6$_{3}mc$
is also shown. Light blue circles indicate experimental pattern and
the red line indicates the theoretical pattern while the green line
is the difference between the two. The Bragg positions are shown as
vertical lines. The (hkl) planes corresponding to the peaks are also
marked in the figure.}
\end{figure}

Fig. \ref{fig:XRD} shows the XRD refinement pattern which yields
R$_{p}=5.7\%$, R$_{wp}=7.7\%$ and goodness of fit (GOF)$=9.70$.
The lattice parameters obtained from the refinement are shown in Table
\ref{tab:LatticeConstant} and compared with the values with Ba$_{3}$RuTi$_{2}$O$_{9}$
(reported earlier), Ba$_{3}$IrTi$_{2}$O$_{9}$ and BaTiO$_{3}$
(parent compound). Values obtained by us are in good agreement with
others. The atomic positions and occupancies resulting from our refinement
are shown in Table \ref{tab:XYZ positions}.

\begin{table}
\caption{\label{tab:LatticeConstant} The lattice constants of Ba$_{3}$RuTi$_{2}$O$_{9}$
obtained from our refinement are compared with literature data.}

\begin{tabular}{|c|c|c|c|c|}
\hline 
Material & Space group & a ($\textrm{\AA}$) & c ($\textrm{\AA}$) & Source\tabularnewline
\hline 
\hline 
Ba$_{3}$RuTi$_{2}$O$_{9}$ & P6$_{3}mc$ & 5.7204(3) & 14.0109(3) & this work\tabularnewline
\hline 
Ba$_{3}$RuTi$_{2}$O$_{9}$ & P6$_{3}mc$ & 5.7056 & 14.0093 & Ref. \cite{Maunders-ActaCrysta-61-2005}\tabularnewline
\hline 
Ba$_{3}$IrTi$_{2}$O$_{9}$ & P6$_{3}mc$ & 5.7214 & 14.0721 & Ref. \cite{Dey-PRB-86-2012}\tabularnewline
\hline 
BaTiO$_{3}$ & P6$_{3}/mmc$ & 5.7238 & 13.9649 & Ref. \cite{Akimoto-ActaCrystC-50-1994}\tabularnewline
\hline 
\end{tabular}
\end{table}

\begin{table}
\centering{}\caption{\label{tab:XYZ positions}The atomic parameters obtained by refining
x-ray powder diffraction data for Ba$_{3}$RuTi$_{2}$O$_{9}$ at
room temperature with the space group P6$_{3}$mc. }
\begin{tabular}{|c|c|c|c|c|c|}
\hline 
\multicolumn{1}{|c}{} &  & x & y & z & g\tabularnewline
\hline 
Ba(1) & 2a & 0 & 0 & 0.249(3) & 1.00\tabularnewline
\hline 
Ba(2) & 2b & 1/3 & 2/3 & 0.085(3) & 1.00\tabularnewline
\hline 
Ba(3) & 2b & 1/3 & 2/3 & 0.397(3) & 1.00\tabularnewline
\hline 
Ru(1) & 2b & 1/3 & 2/3 & 0.651(3) & 0.61(9)\tabularnewline
\hline 
Ti(1) & 2b & 1/3 & 2/3 & 0.651(3) & 0.39(9)\tabularnewline
\hline 
Ti(2) & 2b & 1/3 & 2/3 & 0.839(3) & 0.69(9)\tabularnewline
\hline 
Ru(2) & 2b & 1/3 & 2/3 & 0.839(3) & 0.31(9)\tabularnewline
\hline 
Ti(3) & 2a & 0 & 0 & 0.510(3) & 0.92(4)\tabularnewline
\hline 
Ru(3) & 2a & 0 & 0 & 0.510(3) & 0.08(4)\tabularnewline
\hline 
O(1) & 6c & 0.160(3) & 0.838(3) & 0.575(3) & 1.00\tabularnewline
\hline 
O(2) & 6c & 0.488(4) & 0.511(4) & 0.749(4) & 1.00\tabularnewline
\hline 
O(3) & 6c & 0.160(3) & 0.838(3) & 0.915(3) & 1.00\tabularnewline
\hline 
\end{tabular} 
\end{table}

As a result of site sharing between Ti$^{4+}$ and Ru$^{4+}$ ions,
the Ru$^{4+}$ ($S$ = 1) edge shared triangular plane will be highly
depleted. A probable situation is shown in Fig. \ref{fig:Structure}(iii)
where many of the misplaced Ru$^{4+}$ ions can still manage to interact
with other Ru$^{4+}$ ions forming clusters, while some, called orphan
spins, remain isolated.

\begin{figure}
\centering{}\includegraphics[scale=0.3]{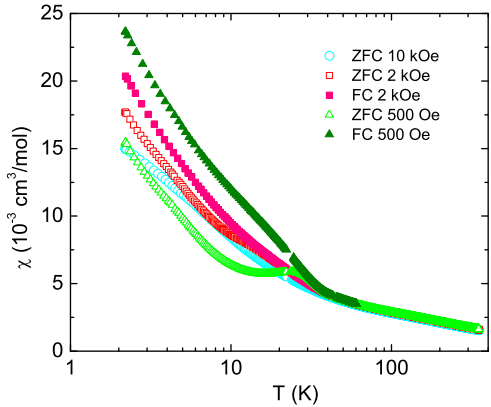}\caption{\label{fig:Chi-T}The magnetic susceptibility of Ba$_{3}$RuTi$_{2}$O$_{9}$
at different fields is shown as a function of temperature in semi-log
scale. Solid symbols represent FC and open symbols represent ZFC measurements. }
\end{figure}

\begin{figure}
\centering{}\includegraphics[scale=0.3]{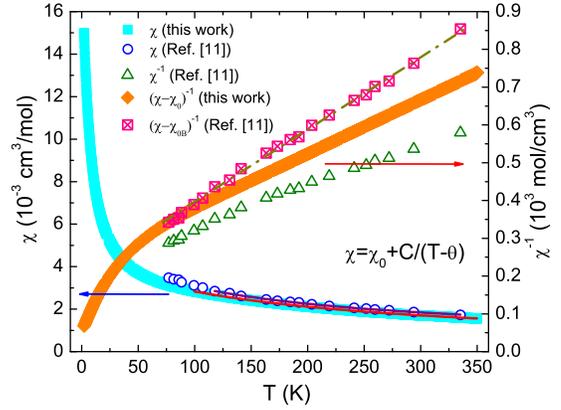}\caption{\label{fig:CWfit} The magnetic susceptibility of Ba$_{3}$RuTi$_{2}$O$_{9}$
and its fitting with the Curie-Weiss law (red solid line) is shown.
The inverse susceptibility (after subtracting $\chi_{0}$) is shown
with orange diamonds. The inverse susceptibility data of Ba$_{3}$RuTi$_{2}$O$_{9}$
obtained from Ref. \cite{Byrne-JSSC-2-1970} is shown as green triangles.
The corresponding susceptibility is shown as blue circles with its
fit with CW law (violet solid line). The inverse susceptibility after
subtracting $\chi_{0\mathrm{B}}$ (obtained from fit, see text) is
also shown with pink squares. The dashed line is a guide to eye. All
the susceptibilities are plotted on the left axis while the inverse
susceptibilities correspond to the right axis. The solid symbols denote
the data from this work and open symbols are used for data from Ref.
\cite{Byrne-JSSC-2-1970}.}
\end{figure}

Further we have measured susceptibility of the material at different
fields as shown in Fig. \ref{fig:Chi-T}. No long range magnetic ordering
down to $2$\,K is found whereas a splitting between ZFC and FC data
at $30$\,K is seen which may be indicative of a transition to a
frozen magnetic state. With increasing field, the splitting becomes
less prominent and vanishes at an applied field of $10$\,kOe. We
have fitted our susceptibility data with the Curie-Weiss (CW) formula
$\chi=\chi_{0}+C/(T-\theta)$ in the temperature range $100-350$\,K
(shown in Fig. \ref{fig:CWfit}) which yields the temperature independent
susceptibility $\chi_{0}=2.05\times10^{-4}$\,cm$^{3}$/mol, Curie
constant $C=0.7$\,cm$^{3}$K/mol and $\theta_{\mathrm{CW}}=-166$\,K.
For insulating oxides one can consider $\chi_{0}=\chi_{core}+\chi_{vv}$,
where $\chi_{core}$ is the core diamagnetic susceptibility and $\chi_{vv}$
is Van-Vleck paramagnetic susceptibility. For our sample $\chi_{core}=\lyxmathsym{\textminus}2.32\times10^{\lyxmathsym{\textminus}4}$\,cm$^{3}$/mol
\cite{Selwood-magnetochemistry}, which results in $\chi_{vv}=4.37\text{\texttimes}10^{\text{\textminus}4}$\,cm$^{3}$/mol.
This value of $\chi_{vv}$ is similar to that found in other Ru based
insulating oxides like La$_{2}$RuO$_{5}$ \cite{Riegg-PRB-2012-La2RuO5}
and La$_{2}$LiRuO$_{6}$ \cite{Aharen-PRB-2009-La2LiRuO6}. Inverse
susceptibility of the material is shown on the right axis of Fig.
\ref{fig:CWfit} (after subtracting the temperature independent part
$\chi_{0}$) which is linear with temperature in a wide range of temperature
and deviates from linearity below $T\sim80$\,K. 

We now compare our susceptibility data with those of Ref. \cite{Byrne-JSSC-2-1970}
by Byrne \textit{et al.} A slope change was found by Byrne \textit{et
al.} in the inverse susceptibility of Ba$_{3}$RuTi$_{2}$O$_{9}$
at $150$\,K which is absent in our data. Moreover, they found $\theta_{\mathrm{CW}}=-256$\,K
and $\mu_{eff}=2.86\mu_{B}$ which are higher than $\theta_{\mathrm{CW}}=-166$\,K
and $\mu_{eff}=2.37\mu_{B}$ obtained from our analysis. To clarify
this discrepancy, we have reanalysed the published data as shown in
Fig. \ref{fig:CWfit}. The green, open triangles represent the inverse
susceptibility data as published in Ref. \cite{Byrne-JSSC-2-1970}.
The corresponding susceptibility is shown by blue circles on the left
axis and fitted by us with the CW law in the range $118-335$\,K.
This fitting gives the temperature independent susceptibility $\chi_{0\mathrm{B}}=5.55\times10^{-4}$\,cm$^{3}$/mol,
$C=0.54$\,cm$^{3}$K/mol ($\mu_{eff}=2.08\mu_{B}$) and $\theta_{\mathrm{CW}}=-116$\,K.
These values are somewhat closer to the values obtained by us. Subtracting
this $\chi_{0\mathrm{B}}$, we have plotted ($\chi-\chi_{0\mathrm{B}}$)$^{-1}$
which is linear in the whole temperature range and the kink at $150$\,K
is also absent. Apparently, Byrne \textit{et al.} did not use the
correct temperature-independent susceptibility, leading them to infer
an incorrect Curie term. Also, they calculated $\mu_{eff}$ based
on a single susceptibility data point (one temperature) which certainly
can mislead.

The $C$ value obtained from the CW fit of our data is almost $70\%$
of the paramagnetic $S=1$ value. Normally in $3d$ transition metal
oxides the $C$ value turns out as expected from a pure spin contribution.
But for $5d$ transition metal oxides the $C$ value obtained is normally
much smaller than expected from a pure spin contribution. In case
of Ba$_{3}$IrTi$_{2}$O$_{9}$ this was only $40\%$ of the expected
spin-only value \cite{Dey-PRB-86-2012}. For $4d$ based systems the
impact of the orbital effect is expected to be in between that in
$3d$ and $5d$ materials. Hence the reduction in $C$ obtained from
our analysis is most likely an effect of spin-orbit coupling present
in the system. A high ($-166$\,K) $\theta_{\mathrm{CW}}$ obtained
from CW fit confirms that the Ru$^{4+}$ spins are strongly correlated.
Probably a large site sharing is responsible in forming some magnetic
clusters which give rise to frozen moments below $30$\,K.

\begin{figure}
\centering{}\includegraphics[scale=0.3]{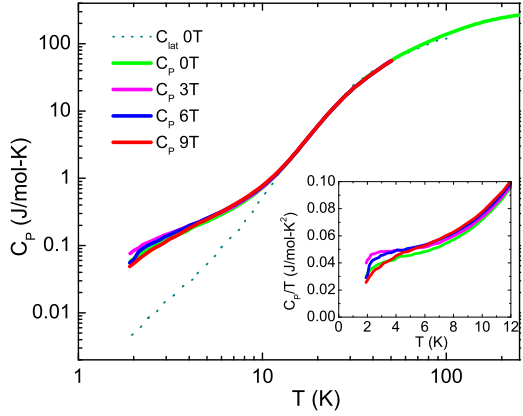}\caption{\label{fig:SpHeat} The heat capacity $C_{\mathrm{P}}$ of Ba$_{3}$RuTi$_{2}$O$_{9}$
measured at different fields is shown in a log-log scale. The lattice
heat capacity (taken from Ref. \cite{Zhou-PRL-106-2011}) is also
shown. Inset: $C_{\mathrm{P}}/T$ is shown as a function of $T$ in
the low-$T$ region.}
\end{figure}

The heat capacity ($C_{\mathrm{P}}$) of a material gives prominent
signatures pertaining to any structural or magnetic transitions. We
have measured $C_{\mathrm{P}}$ at different fields as shown in Fig.
\ref{fig:SpHeat}. \textcolor{black}{No anomaly is found in our $C_{\mathrm{P}}$
measurement in the }\textcolor{black}{\emph{T}}\textcolor{black}{{}
range $1.8\mathrm{K}\leq T\leq295$K.}

The total heat capacity of the sample has three contributions namely,
(i) lattice contribution ($C_{\mathrm{lat}}$) (ii) magnetic contribution
($C_{\mathrm{M}}$) and (iii) Schottky contribution ($C_{\mathrm{Sch}}$)
from orphan spins. The nuclear Schottky anomaly normally appears at
much lower temperature and probably will not affect our data which
is down to $1.8$\,K only. To extract the magnetic heat capacity
we have to subtract out the $C_{\mathrm{lat}}$ and $C_{\mathrm{Sch}}$
from the total heat capacity. We now attempt to estimate the magnetic
contribution by subtracting the other two parts.

The generalised heat capacity expression for two level Schottky anomaly
is \cite{ESRGopal,He-APL-94-2009-Schottky}, 
\begin{equation}
C_{\mathrm{Sch}}(\Delta)=R\frac{g_{0}}{g_{1}}\left(\frac{\Delta}{k_{B}T}\right)^{2}\frac{exp\left(\frac{\Delta}{k_{B}T}\right)}{\left[1+\frac{g_{0}}{g_{1}}exp\left(\frac{\Delta}{k_{B}T}\right)\right]^{2}}\label{eq:SchottkyEq}
\end{equation}

where $\Delta$ is the level splitting, $R$ is the universal gas
constant, $k_{B}$ is the Boltzman constant and $g_{0}$ and $g_{1}$
are the degeneracies of ground state and excited state, respectively.
For spin $S=1$ (Ru$^{4+}$) systems, Eq. \ref{eq:SchottkyEq} can
be modified with the approximation as $g_{0}=1$ and $g_{1}=2$ \cite{ESRGopal}.

\begin{figure}
\centering{}\includegraphics[scale=0.3]{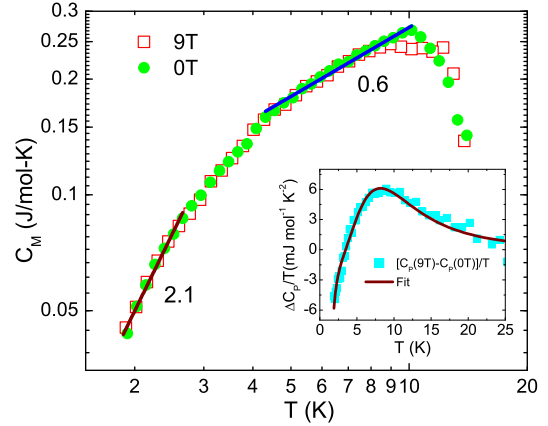}\caption{\label{fig:MagHC} The magnetic heat capacity data of Ba$_{3}$RuTi$_{2}$O$_{9}$
at $0$T and $9$T are shown in a log-log scale. The solid lines show
power law fits with temperature with the power shown in figure. Inset:
{[}$C_{\mathrm{P}}$ ($9$T)- $C_{\mathrm{P}}$ ($0$T){]}/$T$ and
its fit with Schottky model (solid line) is shown.}
\end{figure}

We have subtracted $C_{\mathrm{P}}$($0$T) from $C_{\mathrm{P}}$($9$T)
data and further divided that by temperature. Any lattice contribution
is therefore removed by this procedure. The {[}$C_{\mathrm{P}}$($9$T)-$C_{\mathrm{P}}$($0$T){]}/$T$
data are shown in the inset of Fig. \ref{fig:MagHC} with its fit
to $f[C_{\mathrm{Sch}}(\Delta_{9\mathrm{T}})-C_{\mathrm{Sch}}(\Delta_{0\mathrm{T}})]/T$.
In this expression, $f$ is the fraction of orphan spins present in
the sample and $C_{\mathrm{Sch}}(\Delta_{9\mathrm{T}}),$ and $C_{\mathrm{Sch}}(\Delta_{0\mathrm{T}})$
are the Schottky anomalies corresponding to the level splittings of
$\Delta_{9\mathrm{T}}$ and $\Delta_{0\mathrm{T}}$ with applied magnetic
fields $9$T and $0$T, respectively. The fit is reasonably good and
suggests that the field dependence of $C_{\mathrm{P}}$ is coming
from $C_{\mathrm{Sch}}$ only. From the fit we found $\Delta_{9\mathrm{T}}/k_{B}=27.7$\,K,
$\Delta_{0\mathrm{T}}/k_{B}=1.5$\,K and $\sim1\%$ orphan spins
are present in the system. Putting the corresponding $\Delta$ (obtained
from fitting) in Eq. \ref{eq:SchottkyEq} and multiplying by $f$,
we can get the Schottky contribution ($C_{\mathrm{Sch}}$) to the
heat capacity for a particular field. 

To subtract the lattice contribution ($C_{\mathrm{lat}}$) from the
total heat capacity ($C_{\mathrm{P}}$) one needs a non-magnetic analog
of the system. Unfortunately, we didn't find a good non-magnetic analog
of Ba$_{3}$RuTi$_{2}$O$_{9}$. However we have earlier used Ba$_{3}$ZnSb$_{2}$O$_{9}$
(taken from Ref. \cite{Zhou-PRL-106-2011}) as a non-magnetic analog
in the case of Ba$_{3}$IrTi$_{2}$O$_{9}$ \cite{Dey-PRB-86-2012}.
Ba$_{3}$ZnSb$_{2}$O$_{9}$ can be used as a non magnetic analog
in the present case as well since both have similar crystal structure.
The heat capacities of these two samples have a mismatch at higher
temperature due to the difference in their molecular mass and volume.
We have used a multification factor of $\sim0.67$ so that $C_{P}$
data of Ba$_{3}$ZnSb$_{2}$O$_{9}$ matches with that of Ba$_{3}$RuTi$_{2}$O$_{9}$
in the nominal temperature range $40-60$\,K (shown in Fig. \ref{fig:SpHeat}).
The corrected heat capacity of Ba$_{3}$ZnSb$_{2}$O$_{9}$ can be
considered as the lattice contribution ($C_{\mathrm{lat}}$) to the
total heat capacity of Ba$_{3}$RuTi$_{2}$O$_{9}$. This $C_{\mathrm{lat}}$
is subtracted from $C_{\mathrm{P}}$ and we are left with the $C_{\mathrm{M}}$
and $C_{\mathrm{Sch}}$. Further we have subtracted $C_{\mathrm{Sch}}$
obtained from our previous analysis and finally obtained $C_{\mathrm{M}}$.

The $C_{\mathrm{M}}$ is independent of field as shown in Fig. \ref{fig:MagHC}
and follows a power law in temperature with power $0.6$ in the range
$4.3-10$\,K and with power $2.1$ below $2.7$\,K. This suggests
that spin excitations are present in this system. We have also estimated
the entropy change $\triangle$S$_{\mathrm{M}}$=$0.36$\,J/mol-K
by integrating $C_{\mathrm{M}}/T$. This value is much smaller than
the expected value of $R$$\ln(2S+1)$ = $9.13$\,J/mol-K for a spin-only
system.\textcolor{black}{{} This reduction could be partly due to short-range
magnetic ordering taking place in the system starting at high temperatures
and the entropy is released over a large range of temperature. In
addition, a strong spin-orbit coupling suppressing the magnetic moments
could also be responsible. Whereas other reasons like error in estimation
of $C_{\mathrm{lat}}$ due to lack of a proper non-magnetic analog
may also play a role, it appears that a majority of spins remain disordered
and do not freeze.}

\section{conclusions}

We have prepared Ba$_{3}$RuTi$_{2}$O$_{9}$ (a $4d$ analog of the
$5d$ system Ba$_{3}$IrTi$_{2}$O$_{9}$) and measured its magnetic
susceptibility and heat capacity at different fields. No long range
magnetic ordering is found down to $1.8$\,K despite a large $\theta_{\mathrm{CW}}$
obtained from CW fitting of susceptibility data. This indicates that
the spins are strongly correlated and the system is highly frustrated.
The Curie constant is about $30\%$ smaller than that expected for
isolated $S=1$ moments. This might be because of a spin-orbit coupling
present in this system. Splitting of ZFC and FC susceptibility data
at $30$\,K indicates formation of a frozen magnetic phase. Magnetic
heat capacity of this material is independent of field and approaches
a power law at low temperature with power $2.1$. These measurements
suggest that Ba$_{3}$RuTi$_{2}$O$_{9}$ is a highly frustrated system
but the nature of its ground state is not very clear. Local probe
investigations like $\mu$SR could be useful to check if the spin
freezing happens in the bulk of the sample or not.

\section{acknowledgement}

We acknowledge the financial support from Indo-Swiss Joint Research
Program, Department of Science and Technology, India. We thank K.
Prsa for his help in some supporting measurements and H. M. Ronnow
for useful discussions.

\end{document}